\PassOptionsToPackage{table}{xcolor}
\documentclass[manuscript]{acmart}
\AtBeginDocument{%
  }

\usepackage{float}
\usepackage{graphicx}
\usepackage{amsmath}

\usepackage{soul}
\usepackage{booktabs}
\usepackage{multirow}

\usepackage{siunitx}
\usepackage{url}
\usepackage{hyperref}

\usepackage{xcolor}
\usepackage{subfiles}
\usepackage{algorithm}
\usepackage{algpseudocode}
\usepackage{tcolorbox}
\usepackage{multicol}
\usepackage{textcomp}
\usepackage{subcaption}
\usepackage{caption}
\usepackage{ragged2e}
\usepackage{soul}
\usepackage{hyperref}
\usepackage{multirow}
\usepackage{fancybox}
\usepackage{enumitem}
\usepackage{graphicx}
\usepackage{balance}
\usepackage{color}
\usepackage{float}
\usepackage{tabularx}
\usepackage{adjustbox}
\usepackage{array}
\usepackage{amsmath}
\usepackage{xspace}
\usepackage{url}
\usepackage{tikz}
\usepackage{caption}
\usepackage{pgfplots}
\usepackage{listings}
\usepackage{color}
\usepackage{graphicx}
\definecolor{dkgreen}{rgb}{0,0.6,0}
\definecolor{gray}{rgb}{0.5,0.5,0.5}
\definecolor{mauve}{rgb}{0.58,0,0.82}
\pgfplotsset{compat=1.16}
\usepackage{pgf-pie}
\usetikzlibrary{pgfplots.statistics,calc}
\usepackage{algorithm}
\usepackage{algpseudocode}
\usepackage{subcaption}
\usepackage{listings}

\usepackage{tcolorbox}

\makeatletter
\newcommand{\mybox}[1]{%
	\setbox0=\hbox{#1}%
	\setlength{\@tempdima}{\dimexpr\wd0+13pt}%
	\begin{tcolorbox}[boxrule=0.5pt, colback=white, arc=4pt,
		left=6pt,right=6pt,top=6pt,bottom=6pt,boxsep=0pt]
		#1
	\end{tcolorbox}
}
\usepackage{tikz}
\newcommand*\circled[1]{\tikz[baseline=(char.base)]{
            \node[shape=circle,draw,inner sep=2pt] (char) {#1};}}

\definecolor{codegreen}{rgb}{0,0.6,0}
\definecolor{codegray}{rgb}{0.5,0.5,0.5}
\definecolor{codepurple}{rgb}{0.58,0,0.82}
\definecolor{backcolour}{rgb}{0.95,0.95,0.92}

\lstdefinestyle{mystyle}{
  language=Python,
  aboveskip=3mm,
  showstringspaces=false,
  columns=flexible,
  numbers=none,
  backgroundcolor=\color{backcolour},
  commentstyle=\color{codegreen},
 keywordstyle=\color{magenta},
    numberstyle=\tiny\color{codegray},
    stringstyle=\color{codepurple},
    basicstyle=\small\ttfamily,
    breakatwhitespace=false,         
    breaklines=false,                 
    captionpos=b,                    
    keepspaces=false,                 
    numbersep=5pt,                  
    showspaces=false,                
    showstringspaces=false,
    showtabs=false,                  
    tabsize=2,
    escapeinside=``
}
\definecolor{nima2}{RGB}{1.0, 0.49, 0.0}
\definecolor{songcolor}{RGB}{191,191,191}
\definecolor{color}{RGB}{0.13, 0.67, 0.8}
\definecolor{aruncolor}{RGB}{51,255,51}

\newcommand{\tool}{\texttt{EPiC}}

\usepackage{pgfplots}
\pgfplotsset{width=6cm, compat=1.9}
\graphicspath{ {./figures/} }

\begin{document}

\title{Automated Prompt Engineering for Cost-Effective Code Generation Using Evolutionary Algorithm}
\author{Hamed Taherkhani}
\email{hamedth@yorku.ca}
\orcid{0009-0004-0897-4800}
\affiliation{%
  \institution{York University}
  \city{Toronto}
  \state{Ontario}
  \country{Canada}
}

\author{Melika Sepidband}
\affiliation{%
  \institution{York University}
  \city{Toronto}
  \country{Canada}}
\email{melikasp@yorku.ca}
\orcid{0009-0006-3015-490X}

\author{Hung Viet Pham}
\affiliation{%
  \institution{York University}
  \city{Toronto}
  \country{Canada}}
\email{hvpham@yorku.ca}
\orcid{0000-0003-0861-8326}

\author{Song Wang}
\affiliation{%
  \institution{York University}
  \city{Toronto}
  \country{Canada}}
\email{wangsong@yorku.ca}
\orcid{0000-0003-0617-2877}

\author{Hadi Hemmati}
\affiliation{%
  \institution{York University}
  \city{Toronto}
  \country{Canada}}
\email{hemmati@yorku.ca}
\orcid{0000-0003-0204-9812}




\renewcommand{\shortauthors}{Taherkhani et al.}

\begin{abstract}
\justifying
Large Language Models (LLMs) have seen increasing use in various software development tasks, especially in code generation. The most advanced recent methods attempt to incorporate feedback from code execution into prompts to help guide LLMs in generating correct code in an iterative process. While effective, these methods could be costly due to numerous interactions with the LLM and extensive token usage. 
 To address this issue, we propose an alternative approach named \underline{E}volutionary \underline{P}rompt Eng\underline{i}neering for \underline{C}ode ({\tool}), which leverages a lightweight evolutionary algorithm to refine the original prompts into improved versions that generate high-quality code, with minimal interactions with the LLM. Our evaluation against state-of-the-art (SOTA) LLM-based code generation agents shows that {\tool} not only achieves up to 6\% improvement in pass@k but is also 2–10 times more cost-effective than the baselines.
\end{abstract}


\keywords{Prompt Engineering, Code Generation, Large Language Models, Evolutionary Algorithm}

\maketitle
\section{Introduction}
LLMs have been used in many software development activities such as software testing~\cite{b60,b61}, design~\cite{b62,b63}, requirement engineering~\cite{b64}, code generation~\cite{b25,b26,b27,b28,b29}, maintenance~\cite{b65}, etc.~\cite{b43,b45}. Among these activities, code generation using LLMs has demonstrated significant potential. 

In LLM-based code generation, various prompt engineering techniques, including zero-shot~\cite{b9}, in-context learning~\cite{b46,b47}, RAG (Retrieval Augmented Generation)~\cite{b48}, and task-specific methods~\cite{b49,b50}, have been employed. The most advanced prompt engineering methods for code generation employ various agent-based approaches \cite{b41}. SOTA methods such as Reflexion \cite{b26}, Language Agent Tree Search (LATS) \cite{b27}, AgentCoder \cite{b28},  Large Language Model Debugger (LDB) \cite{b29}, and MetaGPT \cite{b42} are either planning-based or multi-collaborative agents. 
While effective, these methods can be costly due to numerous interactions with LLMs, which result in extensive token usage, making them less attractive in practical settings. 
For instance, as we found in our study (Section~\ref{RQ2}), on the MBPP dataset, LATS requires on average an additional 234k tokens to find the correct implementation of a function that involves multiple interactions with the LLM.

\begin{figure}[t!]
    \centering
    \includegraphics[width=\linewidth]{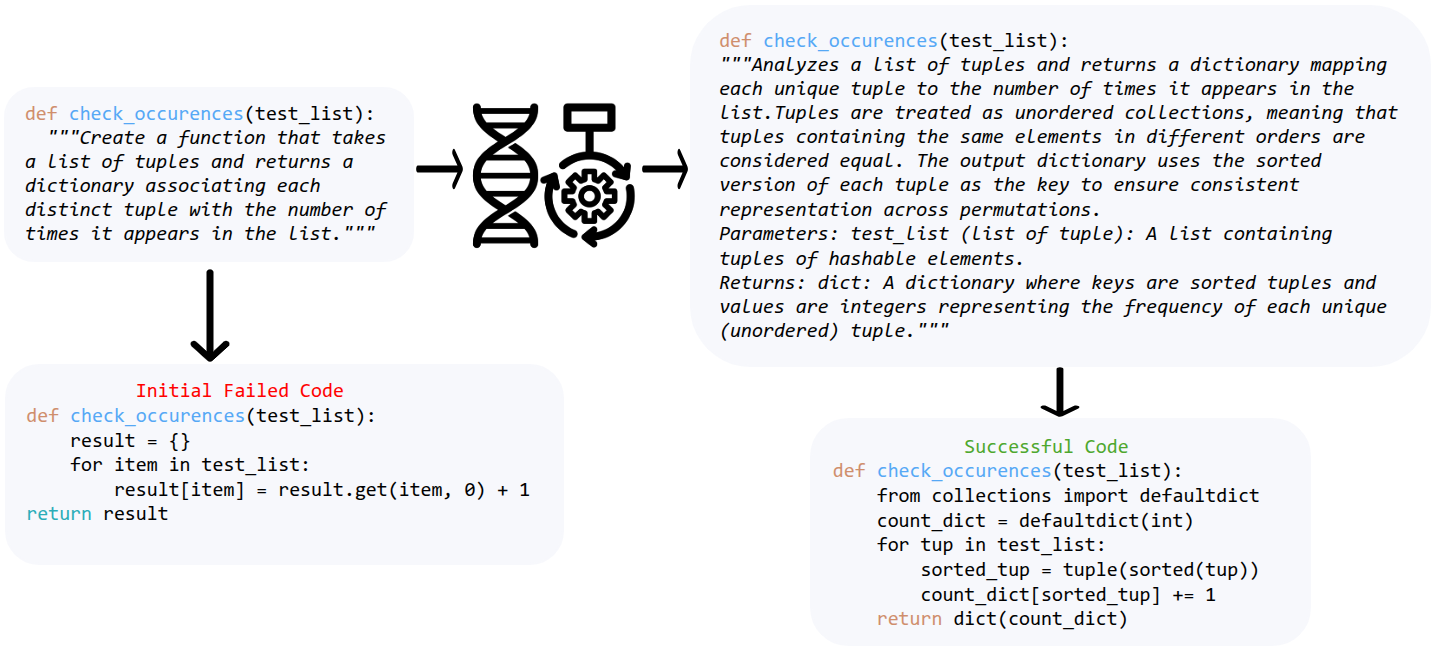}
    \caption{The initial failed prompt (left) and the evolved successful prompt (right)}
    \label{fig:intro}
\end{figure}
One key limitation of these approaches is that the initial prompt fed to the LLM is often suboptimal. If we had an approach that systematically improves the prompt with minimal LLM calls, we could find the optimal prompt. For instance, consider a scenario (Figure \ref{fig:intro}) where a developer’s original prompt is ambiguous. The LLM may repeatedly generate incorrect code until the user (or an agent) clarifies the instructions in subsequent prompts. Each clarification adds new token usage. By contrast, if we systematically evolve the initial prompt and test it, we can converge on correct code more efficiently. 

To accomplish this, we propose Evolutionary Prompt Engineering for Code (EPiC) to refine prompts in a structured and cost-effective way. An Evolutionary Algorithm (EA), in broad terms, maintains a population of candidate solutions (in our case, candidate prompts). It repeatedly evaluates these candidates, selects them based on fitness (test success rate), and mutates them. 
Our approach consists of \textbf{two phases}: Initial Evaluation (\textbf{IE}) and Evolutionary Prompt Engineering (\textbf{EPE}). The first phase involves primary code generation using an initial prompt and its evaluation using a set of test cases. If a correct solution is not achieved, the process moves to the second (EPE) phase, where an initial population of prompts is augmented using the original prompt. For each prompt in the initial population, LLM generates the corresponding code. The generated \textbf{code} is then \textbf{evaluated} against a set of test cases to determine its \textbf{fitness score}. \textbf{Candidate prompts} are selected from the population using a weighted random selection approach and then \textbf{mutated} to form the \textbf{next generation} of prompts. The mutation is carried out using two approaches: one utilizes an LLM guided by a prompt that specifies how to perform the mutation, and the other employs vector embeddings of words to find and replace similar words.


To evaluate the effectiveness of {\tool}, we select three widely used code generation benchmark datasets, i.e., Humaneval+ \cite{liu2023your}, MBPP+ \cite{liu2023your}, and BigCodeBench~\cite{zhuo2024bigcodebench}. 
We implement {\tool} using three SOTA LLMs: OpenAI's o3-mini, Deepseek-v3, and Sonnet 3.7. We also compare {\tool} against three SOTA LLM-based agentic code generation tools, i.e., Reflexion \cite{b26}, LATS \cite{b27}, and LDB \cite{b29}.

We use the $pass@k$ metric, as introduced in \cite{b12}, for accuracy (effectiveness), which estimates the probability that at least one of the top k generated code samples is correct. To evaluate the cost-effectiveness, we introduce a new metric called Additional Token Usage per Solved Problem (ATSP in Section ~\ref{metrics}), which quantifies the additional tokens consumed for each extra problem solved relative to a baseline prompting method, where lower values indicate better cost-effectiveness. Our experiments demonstrate that our approach not only consumes fewer tokens per additional solved problem, resulting in lower ATSP scores, but also outperforms the baseline in terms of pass@k by up to 6\%. In addition, in terms of ATSP, {\tool} is twice as cost-effective as the most cost-effective baseline, Reflexion~\cite{b26}, and almost 10 times more cost-effective than LATS~\cite{b27}.

The main contributions of this paper are as follows:

\begin{enumerate}
    \item To the best of our knowledge, we are the first to explore the code generation task from the cost-effectiveness perspective. 
    \item We propose a novel framework {\tool} that leverages a lightweight evolutionary algorithm to evolve the original prompts toward better ones that produce high-quality code.
 
    \item We have demonstrated the cost-effectiveness of {\tool} in code generation compared to baseline methods.
    \item  We release the data and source code of our experiments to enable other researchers to replicate and extend our study (\href{https://github.com/HamedTaherkhani/EPiC}{https://github.com/HamedTaherkhani/EPiC}). 
\end{enumerate}

\section{Background and Related Work}
\label{relatedWork}

In this section, we briefly explain prompt engineering for LLMs in general and report the most related work in the context of prompt engineering of LLM for code. In the end, we discuss evolutionary algorithm in general and how it aligns with {\tool}.

\begin{figure*}[t!]
\includegraphics[width=\linewidth]{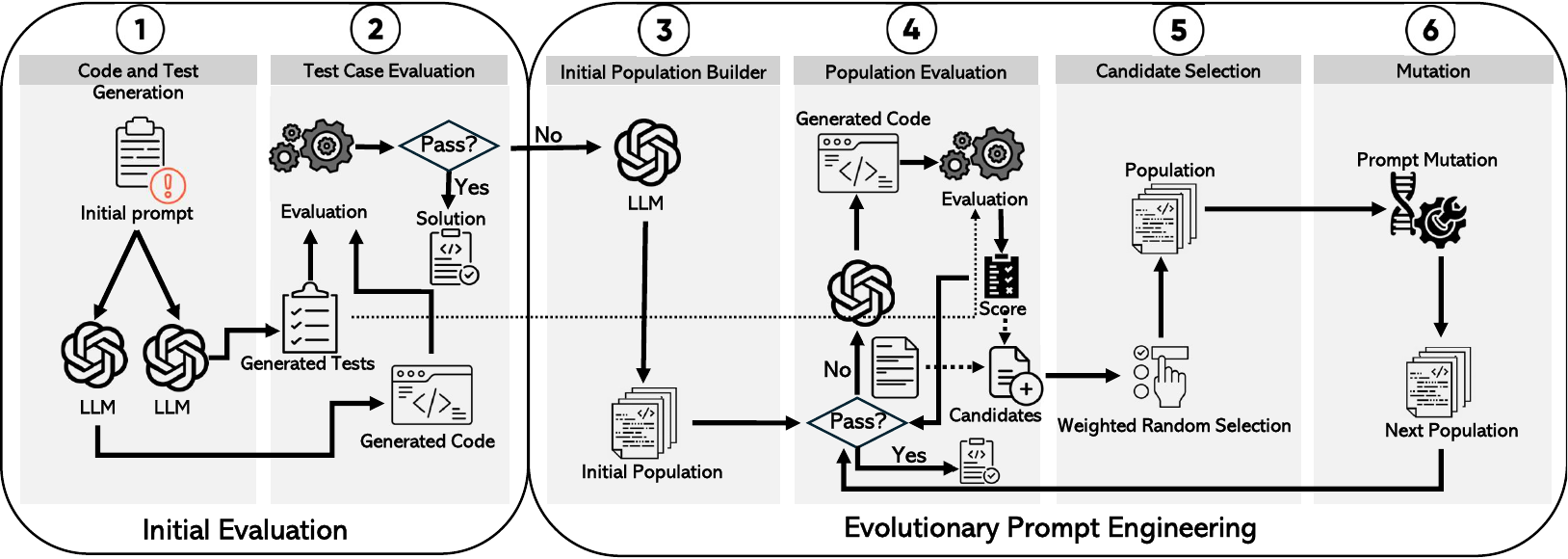}
\caption{Diagram of {\tool}. On the left, the initial evaluation for the initial prompt assessment is depicted, and on the right, the evolutionary process of {\tool} is illustrated.}

\label{fig1}
\end{figure*}

\subsection{Prompt Engineering for LLMs}
A prompt is an input or query provided to a model to generate a response or perform a task. 
Prompt engineering refers to the process of designing and refining prompts to achieve desired outcomes when using LLMs\footnote{https://platform.openai.com/docs/guides/prompt-engineering}.  




There are multiple categories of prompt engineering, including approaches without training, reasoning and logic, reducing hallucination, and evolutionary-based methods \cite{b51}. Zero-shot \cite{b9} and few-shot \cite{b46} prompting fall under the category of approaches without training. Techniques such as chain-of-thought (CoT) prompting \cite{b17}, Automatic Chain-of-Thought (Auto-CoT) \cite{b53}, Self-Consistency \cite{b54}, and knowledge prompting \cite{b18} exemplify reasoning and logic-based methods. To reduce hallucination for prompt engineering, techniques such as Retrieval Augmented Generation (RAG) \cite{b48}, ReAct Prompting \cite{b56}, and Chain-of-Knowledge (CoK) Prompting \cite{b57} are employed. Evolutionary algorithms are utilized to optimize prompts, as demonstrated by EvoPrompt \cite{b22} and PromptBreeder \cite{b37}. There are other solutions, such as Automated Prompt Engineering~\cite{b19} and Prompt Optimization with Textual Gradients (ProTeGi)~\cite{b21}.


Zero-shot prompting~\cite{b9} enables task execution without training data by using well-designed prompts, relying on the model’s pre-existing knowledge. Few-shot prompting~\cite{b46} provides limited examples to guide understanding but requires more tokens, making it less practical for long texts and susceptible to bias from example selection. CoT~\cite{b17} improves LLM reasoning by integrating intermediate steps and breaking down complex tasks to improve arithmetic, common sense, and symbolic reasoning, especially in larger models. Auto-CoT~\cite{b53} automates reasoning step generation using diversity-based sampling to create multiple reasoning paths. Automated prompt engineering leverages agents to refine prompts based on LLM feedback. Zhou et al.~\cite{b19} introduced APE, which generates and selects instructions by optimizing a scoring function. Pryzan et al.~\cite{b21} proposed ProTeGi, a non-parametric method using natural language gradients and beam search to enhance prompt editing, improving efficiency across NLP tasks. Evolutionary algorithms have been used for LLM prompt engineering in NLP tasks. EvoPrompt \cite{b22} automates this by iteratively refining prompts using mutation and crossover. Similarly, PromptBreeder \cite{b37} employs task and mutation prompts for iterative improvement. Similar approaches for prompt enhancement using LLMs through EAs are also proposed in \cite{b38} and \cite{b39}.


In contrast to existing evolutionary prompt engineering techniques, {\tool} is particularly tailored for prompt engineering in coding tasks, with a fitness function defined based on the pass rate of test cases. In addition, it focuses on \textbf{cost-effectiveness} throughout the process. To achieve this, it minimizes the calls to external LLMs and implements the mutation operator using local lightweight \textbf{word embeddings} libraries. 
\subsection{Prompt Engineering of LLMs for Code Generation}


DyLAN \cite{b25} enhances LLM performance by dynamically assembling agent teams based on task queries, unlike fixed-agent methods. It improves efficiency with inference-time agent selection, early stopping, and an unsupervised Agent Importance Score for optimizing agent selection. 
Reflexion \cite{b26} is a reinforcement-based framework in which language agents learn from linguistic feedback. By reflecting on task feedback and storing reflective text in memory, agents improve decision-making in future trials. 
LATS \cite{b27} (Language Agent Tree Search) is a framework that integrates the planning, acting, and reasoning abilities, inspired by Monte Carlo tree search. It utilizes LLMs as agents, value functions, and optimizers while incorporating external feedback for more adaptive problem-solving. 
AgentCoder \cite{b28} employs a multi-agent framework with specialized roles: the programmer generates and refines code, the test designer creates test cases, and the test executor runs tests and provides feedback. This collaborative approach enhances code generation efficiency, surpassing single-agent models. 
Zhong et al. \cite{b29} introduced Large Language Model Debugger (LDB), a framework that segments programs into basic blocks and tracks intermediate variable values during runtime. LDB enables LLMs to focus on simpler code units and verify their correctness block by block.

In contrast to existing methods and to the best of our knowledge, {\tool} is the first evolutionary-based prompt engineering method for code generation. It employs a lightweight process to identify the optimal solution in a cost-effective manner. To achieve this, {\tool} utilizes a local embedding function to implement mutation operators on text to reduce the cost of iterative prompt engineering for code generation. It also guides the search over iterations using the fitness function in Section \ref{obj-func}, which helps in finding the best prompts.

\subsection{Evolutionary Algorithms}

Evolutionary algorithms (EA) are population-based metaheuristics inspired by natural selection. One key advantage of EA is that it can optimize using only a fitness function without requiring gradient information, making it particularly suitable for problems where gradients are difficult or impossible to compute, such as prompt optimization, where the pass rate of generated code is a discrete and non-differentiable function. EA typically follows a cycle of initialization, evaluation, selection, variation, and iteration. The process begins with an initial population of candidate solutions, which, in {\tool}, are variations of the original prompt. Each candidate is evaluated using a fitness function that measures quality based on the pass rate of test cases. Selection then occurs through a weighted random strategy that prioritizes prompts demonstrating higher pass rate. Variation operators, such as mutation, introduce modifications to candidates, and this iterative cycle continues until convergence or a predefined stopping criterion is met.

While EA is effective in navigating complex search spaces, it also has drawbacks, such as potential inefficiency when fitness evaluations are expensive and the risk of premature convergence. However, for optimizing prompts—where evaluating fitness is relatively straightforward but computing gradient information is non-trivial—EA is a natural and effective choice. Over time, this process refines the prompt by focusing on those that lead to code passing more test cases, ultimately producing an optimized prompt whose generated code passes all test cases.  




\section{Evolutionary Prompt Engineering for Code }


\begin{algorithm}[ht]
\footnotesize
\caption{Evolutionary algorithm}
\label{alg:one}
\begin{algorithmic}[1]
    \footnotesize
  \Procedure{EvoALG}{$prompt, tests, popSize$}
  \State $evaluations \gets []$
  \State $population \gets []$
  \State $code, fitness \gets evaluatePrompt(prompt, tests)$
        \If{$fitness$ is $1$}
            \State $return$ $code$    
        \Else
        \State $population \gets generateFirstPopulation(popSize, prompt)$
        \EndIf
        \While{not $convergecriteriaMet(evaluations)$}
            \State $candidates \gets []$
            \For{$indPrompt$ in $population$}
                \State $code, fitness \gets evaluatePrompt(indPrompt, tests)$
                \If{$fitness$ is $1$}
                    \State $return$ $code$
                \Else{}
                    \State $candiates.append([indPrompt,code,fitness])$
                \EndIf
            \EndFor
            \State $evaluations.append(candiates)$
            \State $mutatedPrompts \gets [ ]$ 
            \State $muCandidates \gets chooseCandidates(candidates,N-1)$
            \For{$candid$ in $muCandidates$} 
                \State $mutatedPrompts.append(mutate(candid))$
            \EndFor
            \State $elitePrompt \gets chooseCandidates(candidates,1)$ 
            \State $population \gets elitePrompt + mutatedPrompts$   
        \EndWhile
        \State $return$ $chooseBestCandidate(candidates)$
    \EndProcedure
    \end{algorithmic}
\end{algorithm}

\begin{algorithm}[t]
\footnotesize
    \caption{Mutation process in $sim\_words\_as\_mutator$}
    \label{alg:two}
    \begin{algorithmic}[1]
        \Procedure{mutate}{$prompt, sim\_t,num\_t,
        mutation\_probability$}
        \State $sentence \gets extract\_description(prompt)$
        \State $words \gets sentence.split()$
        \State $sent\_comb \gets []$
        \For{$word$ in $words$}
            \If{$word$ is $proper\_noun$ or $stop\_word$}
                \State $sent\_comb.append([word])$
            \Else
                \State $word\_lem \gets lemmatize(word)$
                \State $sent\_comb.append(get\_related\_words(word\_lem$,
                $sim\_t, num\_t))$
            \EndIf
        \EndFor
        \State $alt\_sentence  \gets " "$
        \For{$j$ in $sent\_comb$}
            \If{$random.random() > mutation\_probability$}\Comment{Use original word}
                \State $alt\_sentence \gets alt\_sentence + j[0]$
            \Else \Comment{Use related word}
                \State $alt\_sentence \gets alt\_sentence + random.choices(j[1:], k=1)$   
            \EndIf
        \EndFor
        \State return $alt\_sentence$
        \EndProcedure
    \end{algorithmic}
\end{algorithm}

This section provides a detailed elaboration of {\tool}. 
As depicted in Figure \ref{fig1}, our approach comprises two phases, i.e., the Initial Evaluation (\textbf{IE}) phase and the Evolutionary Prompt Engineering (\textbf{EPE}) phase. The process begins with a user providing an initial prompt that describes the intended functionality of the code to be generated. In the \textbf{IE} phase, we evaluate the prompt by generating the code based on the original prompt using an LLM. This evaluation determines whether the prompt is sufficient to generate the correct implementation or if it requires a further process in the \textbf{EPE} phase. {\tool}'s algorithm is presented in Algorithm \ref{alg:one}, which will be elaborated upon in the subsequent sections. \textbf{Lines 2 to 6} in Algorithm \ref{alg:one} correspond to the \textbf{IE} phase of {\tool} and \textbf{lines 7 to 30} are the implementaion of the \textbf{EPE} phase.

\subsection{Initial Evaluation}
Initial evaluation (IE) is intended for an early stop in the code generation process if a correct solution is identified. In Step \circled{1} of the \textbf{IE} phase, the initial prompts for code and test case generation are provided to the LLM  (the choice of LLM to be used in our case will be discussed in the Section \ref{experiments}) to generate the corresponding code and test cases. In Step \circled{2}, the generated code is evaluated against the generated test cases. \textit{evaluatePrompt} in \textbf{Line 4} of Algorithm \ref{alg:one} is responsible for generating the code and evaluating the test cases. If any test case fails on the code, we continue with the EPE phase. If all test cases pass, we report the generated code as the final answer.

Note that test cases can be provided in various ways. One approach is to use developer-provided test cases for evaluation, while another is to generate test cases using the LLM. The challenge with developer-provided test cases is that, typically, users do not have test cases before implementation, making the approach impractical for a code generation task. On the other hand, the advantage of using LLM-generated test cases is that it makes the process fully automated. This method is currently used in most SOTA code generation tools \cite{b26}, \cite{b27}, \cite{b28}, \cite{b24}, and \cite{b42}, as part of their internal evaluation. Consequently, we also opted to use LLMs for test case generation to ensure a fully automated approach, assuming no developer-provided test cases. To ensure the functional correctness of these test cases, we validated them by parsing their Abstract Syntax Trees (AST). A test case is considered valid if it successfully parses into a syntactically correct AST. This approach is the same approach used in Reflexion \cite{b26}.

The prompts for code and test case generation that are used in our experiments are provided below. These prompts can be modified if necessary, depending on the dataset.

\mybox{
Test Case Generation Prompt\\
\footnotesize
You are an AI coding assistant who can write unique, diverse, and intuitive unit tests for functions, given the signature and docstring. Use step-by-step reasoning to validate each test case separately to ensure that all of them have the correct expected output. Write the final validated tests inside \$\$\$\$ tags.

Examples:\\
def add3Numbers(x, y, z):\\
    """ Add three numbers together. This function takes three numbers as input and returns the sum of the three numbers."""\\
\$\$\$\$\\
assert add3Numbers(1, 2, 3) == 6\\
assert add3Numbers(-1, 2, 3) == 4\\
assert add3Numbers(1, -2, 3) == 2\\
assert add3Numbers(1, 2, -3) == 0\\
assert add3Numbers(-3, -2, -1) == -6\\
assert add3Numbers(0, 0, 0) == 0\\
\$\$\$\$
}

\mybox{
Code Generation Prompt\\
\footnotesize

You are a Python developer who implements the correct code based on the function description provided. You are given one or more functions to implement. Do not delete import statements from the code snippet. Use at most 1000 words.
}

\subsection{Evolutionary Prompt Engineering}
Evolutionary Prompt Engineering (EPE) is a systematic approach that leverages EA to iteratively refine and optimize prompts for code generation. In the \textbf{EPE}  phase, the first step (Step \circled{3}) involves populating the first generation of individuals, where each individual is a distinct prompt to generate the same code. In Step \circled{3}, we generate multiple prompts by modifying the initial prompt using an LLM agent. Generating this prompt population forms an important part of our evolutionary algorithm. A more detailed description of this process will be provided in Section \ref{part4.1}. This step corresponds to \textbf{line 8} of Algorithm \ref{alg:one}.

In Step \circled{4} of \textbf{EPE}, which corresponds to \textbf{lines 12 to 20} in Algorithm \ref{alg:one}, we generate code for each prompt and evaluate each generated code sample using the same test cases. We define a fitness function based on the ratio of test cases passed. If any of the prompts achieves the maximum fitness score, the process stops. If not, the process continues until all prompts are validated and given a fitness score, which will guide the selection process. The fitness function is the passing rate of each prompt. The mathematical formula is explained in Section \ref{obj-func}. 

 This fitness function is used to select the candidate prompts for mutation in Step \circled{5}. This is implemented in \textit{chooseCandidate} function in Algorithm \ref{alg:one} (\textbf{Lines 22 and 26}), where a weighted random selection algorithm selects candidates randomly, with the probability of selection being proportional to their respective fitness score (Section \ref{obj-func}). This selection occurs with substitution, allowing the same prompt to be chosen multiple times. In Step \circled{5}, we select $N-1$ candidates for mutation and one prompt for ``elitism''. 


Following prompt selection, in Step \circled{6}, {\tool} randomly mutates the selected prompt candidates to better explore the search space of potential prompts in the next generation. The prompt mutation which is detailed in Algorithm \ref{alg:two}, is elaborated in Section \ref{EA}. The next generation is formed by adding the elite prompt to the pool of mutated prompts. \textbf{Lines 21 to 27} in Algorithm \ref{alg:one} correspond to Steps \circled{5} and \circled{6}.


The process of \textbf{EPE} is repeated iteratively until a solution achieves maximum fitness score or until predefined stopping criteria, i.e., reaching the maximum number of iterations or observing no improvement in the fitness scores, are met. In such cases, the best-generated code, determined by its fitness, is chosen and returned (\textbf{line 29} of Algorithm \ref{alg:one}). 

In the next two subsections, we will explain the details of Step \circled{3} (Initial Population Builder) and Step \circled{6} (Prompt Mutation).

\begin{figure*}[t]
  \centering
  \begin{subfigure}[b]{0.32\textwidth}
    \includegraphics[width=\textwidth]{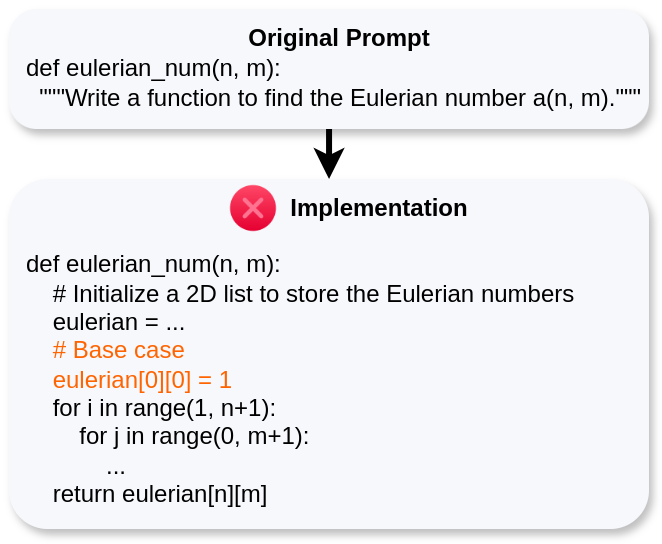}
    \caption{Original prompt and its result}
    \label{fig:f2}
  \end{subfigure}
  \begin{subfigure}[b]{0.67\textwidth}
      \includegraphics[width=\textwidth]{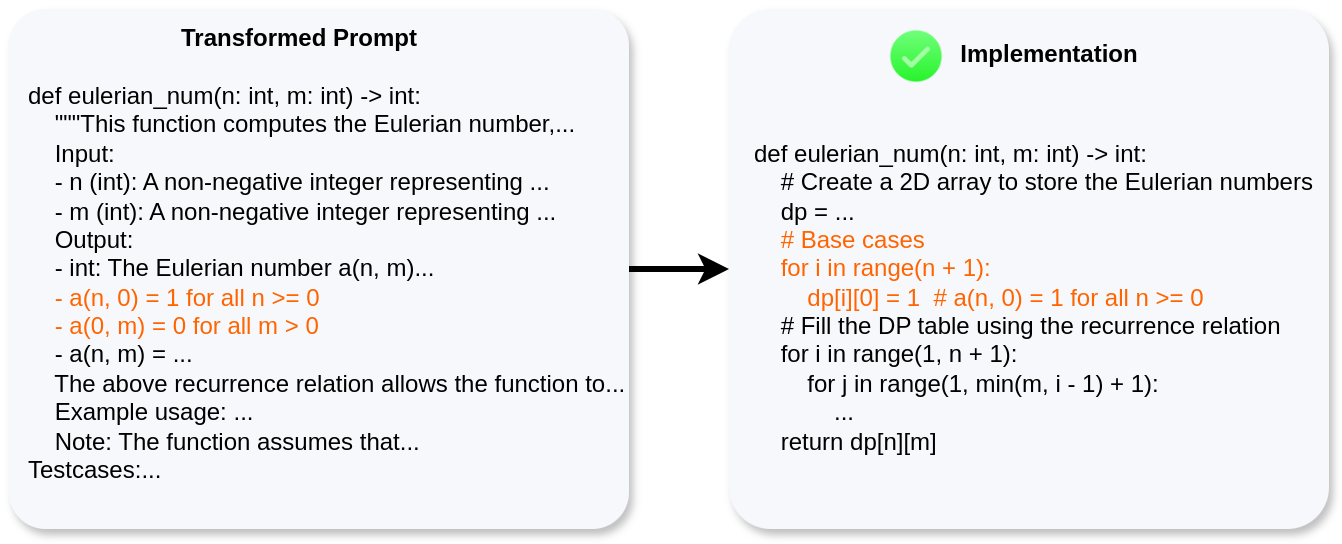}
      \caption{Transformed prompt and its result}
      \label{fig:f1}
  \end{subfigure}
  \caption{An unsuccessful original prompt on the left and the transformed successful version of it on the right. The main issue with the original implementation is the incorrect initialization of the Eulerian number. The orange text in both figures highlights the code and text related to the initialization of the function.}
  \label{fig:8}
\end{figure*}

\subsubsection{Initial Population Builder}
\label{part4.1}
This subsection provides a detailed explanation of Step \circled{3} in Figure \ref{fig1}. Human-written prompts are often very brief and lack the necessary information or context. This results in insufficient descriptions and a failure to adhere to a well-defined format~\cite{b58}. Thus, manual prompting typically results in several back-and-forth interactions with LLM until all necessary details are given~\cite{b41}. 
Providing the initial prompt in a structured and elaborate format will guide the LLM to generate better results with fewer interactions~\cite{b44}.  

In our evolutionary algorithm (Algorithm~\ref{alg:one}), we require an initial population of prompts. We utilize the LLM to augment multiple prompts based on the initial given prompt. The prompt we use is as follows:
\vspace{-0.1in}
\mybox{
  Please rewrite the function description following the instructions:\\
1- Add input and output types of the function to the description.

2- Elaborate the description so that it is understandable for large language models.

3- Keep the original test cases and add three test cases to the description to cover the edge cases. Do not separate the generated test cases and the original ones.

4- Keep the structure of the function and add the description as a comment in the function. Use at most 500 words. 
}

We employ a high-temperature setting (0.6) for the LLM to stimulate creativity in prompt generation. 
Elaborate prompts, which include explicit input-output types and test cases, more effectively direct the language model to produce the desired behavior compared to the original prompts. 
Figures \ref{fig:f2} and \ref{fig:f1} show an example of such an elaborate prompt. 

Figure \ref{fig:f2} illustrates the original prompt provided to an LLM, i.e., GPT-4o, which results in an implementation that fails to pass the test cases. Conversely, Figure \ref{fig:f1} depicts a transformed version of the prompt that successfully guides the LLM toward the correct implementation. The primary issue with the original implementation is the incorrect initialization of the Eulerian number. In contrast, the transformed prompt includes specific instructions on how to properly initialize the Eulerian number, leading to a correct implementation. The orange text in both figures highlights the code and text related to the initialization.

After generating the elaborate prompt, we add the previously generated test cases (Step \circled{1} in Figure \ref{fig1}) to each prompt. These test cases will also be used during the evolutionary algorithm's fitness evaluation in Step \circled{4}.

\begin{figure*}[t]
    \centering
    \includegraphics[width=\linewidth]{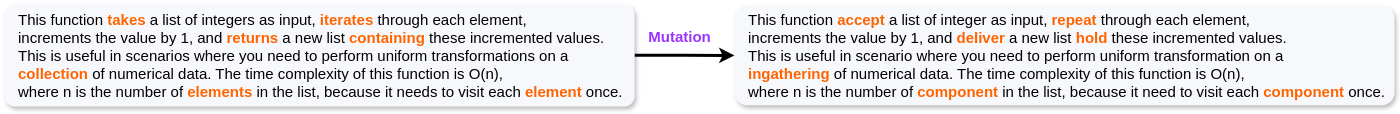}
    \caption{\small{An example of a mutation generated by $sim\_words\_as\_mutator$. The orange words highlight the selected and mutated words in this example.}}
    \label{fig:mutation}
\end{figure*}

\subsubsection{Prompt Mutation}
\label{EA}




This subsection provides a detailed explanation of Step \circled{6} in Figure \ref{fig1}. We employed two kinds of prompt mutation approaches. In the first approach, {\tool} uses an LLM to facilitate the mutation process ($LLM\_as\_mutator$). This approach is inspired by a framework named EvoPrompt~\cite{b22}. In $LLM\_as\_mutator$, we provide LLMs with predefined instructions on how to implement the prompt mutation. 
In the second approach, we employ Natural Language Processing (NLP) techniques using the NLTK and Gensim libraries to alter the prompts by substituting words with their synonyms ($sim\_words\_as\_mutator$). Synonym replacement, when applied with controlled probability, can improve model performance without altering the fundamental semantics of a sentence~\cite{wei-zou-2019-eda}. We extend this concept by integrating $sim\_words\_as\_mutator$ into our prompt evolution framework, reducing computational cost while making meaningful mutations. Since the evolutionary algorithm involves iterative prompt refinement, the cost-effectiveness and linguistic validity of this approach make it an ideal mutation operator. 
This prompt mutation algorithm ($sim\_words\_as\_mutator$) is described in Algorithm \ref{alg:two}. We will compare the performance of both approaches in Section \ref{RQ3}. We use the following prompt for the mutation process in $LLM\_as\_mutator$:
\mybox{
You are a mutation tool. This is a Python function and its description. Please change the description by enhancing its clarity and comprehensibility for sophisticated language models.\\
Please put the changed description between \#Explanation and \#End. Use at most 600 words.
}

In the $sim\_words\_as\_mutator$ (Algorithm~\ref{alg:two}), we first extract the description part of the prompt. For each word in the sentence, if the word is not a stop word or proper noun, it is lemmatized, and the \textit{get\_related\_words()} function (\textbf{line 10} in Algorithm \ref{alg:two}) is called to get its related words. To obtain the related words for a given $word$, we use $WordNet$ and $gensim$'s $GloVe vectors$ ($fasttext-wiki-news-subwords-300$) to calculate the cosine similarity between the word and other related words. We rank the related words to create a list of the most similar words to the original word, and based on the specified $num\_t$ or $sim\_t$, we filter out the least related words and keep words that pass the given threshold. Then, we randomly substitute the word with one of the related words based on a $mutation\_probability$. 
Figure \ref{fig:mutation} illustrates an example of mutation performed by $sim\_words\_as\_mutator$. In this example, selected words are highlighted and replaced with their related synonyms by this algorithm. For instance, the word ``takes'' is randomly chosen for mutation. The algorithm first identifies similar words for this word, resulting in the following list: [``take'', ``make'', ``require'', ``have'', ``carry'', ``get", ``bring", ``accept'', ``lead'', ``hold"]. The algorithm then randomly selects the word ``accept'' and replaces it with ``take''. This process is similarly applied to other randomly selected words, such as ``iterates'', ``returns'', ``containing'', ``collection'', and ``elements''.



\section{Experiment Design}
\label{experiments}
\subsection{Research Questions}
We define the following research questions to evaluate the performance of {\tool}:


\noindent \textbf{RQ1: How does {\tool} perform across different SOTA LLMs and datasets in terms of pass@1 and token usage?} \textit{Motivation: Evaluating {\tool} on the SOTA LLMs provides us with insights into their performance and token usage in code generation. This evaluation also helps us find the best-performing LLM for the next RQ. In addition, it also shows that {\tool}'s performance is not limited to one LLM.}

\noindent \textbf{RQ2: How does {\tool} compare to other iterative-based agents (Reflexion, LATS, and LDB) in terms of pass@1, token usage, and cost-effectiveness across different benchmarks?} \textit{Motivation: cost and pass@1 are the two major factors in identifying the most effective agent. In this RQ, we analyze the cost-effectiveness of the SOTA code generation agents and compare them with {\tool}.}

        





\subsection{Datasets and Models}

We used three datasets for our experiments: HumanEval+ \cite{liu2023your}, MBPP+ \cite{liu2023your}, and BigCodeBenchHard~\cite{zhuo2024bigcodebench}.

\textbf{HumanEval+}~\cite{b12} is a collection of 164 programming problems designed to evaluate the functional correctness of code generated by AI models. Each problem includes a function signature, a detailed description of the task (docstring), the function body, and multiple unit tests to verify the solution. 
HumanEval+ extends the original \textit{HumanEval} dataset by increasing the number of test cases by 80 times using both LLM-based and mutation-based automated test generation techniques. This augmentation helps uncover incorrect LLM-generated code that was previously misclassified as correct due to inadequate testing, leading to more accurate pass@k performance evaluations for LLMs.

\textbf{MBPP+}~\cite{liu2023your} builds upon the MBPP (Mostly Basic Python Programming) dataset, which consists of 974 crowd-sourced programming problems, each with a problem statement, function signature, and three test cases. MBPP+ enhances this dataset by incorporating additional, automatically generated test cases to improve the robustness of functional correctness evaluations for LLMs. Similar to HumanEval+, this augmentation ensures that LLM-generated code is tested against a wider variety of scenarios, including edge cases, leading to more reliable performance assessments. 

\textbf{BigCodeBenchHard}~\cite{zhuo2024bigcodebench} is a challenging subset of BigCodeBench designed to evaluate LLMs on complex programming tasks requiring diverse function calls, compositional reasoning, and precise instruction following. Covering 139 libraries across seven domains, it includes rigorous execution-based evaluation with an average of 5.6 test cases per task. Compared to benchmarks such as HumanEval, it presents longer prompts, higher cyclomatic complexity, and broader tool use.

We employed three SOTA LLMs from different vendors in our study: OpenAI's O3-mini\footnote{https://openai.com/index/openai-o3-mini/}, Antropic's Claude 3.7 Sonnet\footnote{https://www.anthropic.com/claude/sonnet}, and Deepseek V3\footnote{https://github.com/deepseek-ai/DeepSeek-V3}.
\subsection{Evaluation Metrics}
\label{metrics}
\subsubsection{pass@k}
To evaluate the quality of the code generated by LLMs, we utilized the pass@k metric. This metric, introduced in \cite{b12}, is advantageous over metrics such as CodeBLEU and ROUGE because it evaluates code behavior rather than text semantics, which are the focus of the latter metrics. The formula for pass@k is presented in Equation \ref{eq.1}. In this equation, \(E\) denotes the expected value over the problems, \(n\) is the total number of generated samples per task (code samples), and \(c\) is the number of correct samples. For a single problem, this metric estimates the probability that at least one of the top \(k\) samples is correct. When applied to multiple problems, it evaluates the expectation across all problems.

\begin{equation}
pass@k:=\underset{\text { Problems }}{\mathbb{E}}\left[1-\frac{\left(\begin{array}{c}
n-c \\
k
\end{array}\right)}{\left(\begin{array}{c}
n \\
k
\end{array}\right)}\right]
\label{eq.1}
\end{equation}
\subsubsection{ATSP}
To measure the cost-effectiveness, we compare the token usage of a code-generation agent against a baseline approach. Specifically, we define a new metric, \textbf{Additional Token usage per Solved Problem (ATSP)}, which quantifies the additional token consumption required to solve more problems compared to the baseline. For the baseline, we use a zero-shot prompting approach. Given a code generation agent~\( m \) and a baseline method~\( b \), the ATSP metric is formally defined as:
\[
\text{ATSP}(m, b) = \frac{T_m - T_b}{\left(P_m - P_b\right) \times N}
\]
where:
\begin{itemize}
    \item \( T_m, T_b \): Total token usage for method~\( m \) and baseline~\( b \), respectively
    \item \( P_m, P_b \): \text{pass@1} scores (fraction of problems solved) for method~\( m \) and baseline~\( b \), respectively
    \item \( N \): Total number of evaluation problems
\end{itemize}
The numerator (\( T_m - T_b \)) represents the token cost difference, while the denominator (\( (P_m - P_b) \times N \)) reflects the absolute increase in solved problems. \textbf{Lower ATSP} values indicate \textbf{better cost-effectiveness}, as fewer additional tokens are consumed per additional solved problem compared to the baseline. \textit{Note}: ATSP is only meaningful when \( P_m > P_b \); methods failing to outperform the baseline are trivially less cost-effective.
\subsection{Objective Function}
We use \(\mathcal{X}\) and \(\mathcal{Z}\) to denote the prompt space and output responses respectively. Let $S: \mathcal{X} \rightarrow \mathcal{Z}$ be the LLM that takes an input $x \in \mathcal{X}$ and outputs a response $z \in \mathcal{Z}$. The output response is the code implementation of the given prompt. Let's say we have $n$ number of test cases. Consider $\mathcal{T}$ as a set of test case functions. A test case $t_i \in \mathcal{T}$ is a function that maps any output response $z$ to either $0$ or $1$. It can be formally defined as $t_i: \mathcal{Z} \rightarrow \{0,1\}$. The fitness function which is denoted as $\mathcal{F}: \mathcal{X}, \mathcal{T} \rightarrow [0,1]$, maps any input prompt to a value between $0$ and $1$. The fitness function is defined as follows:

\begin{equation}
\label{formula1}
    \mathcal{F}(x, \mathcal{T}) = \frac{\sum_{i=1}^{i=n}t_i(\mathcal{S}(x))}{n}
\end{equation}

The goal of the {\tool} framework is to optimize the fitness function \(\mathcal{F}\) based on the provided test cases \(\mathcal{T}\) by identifying the optimal input \(x\) within the prompt space \(\mathcal{X}\).

\label{obj-func}
\subsection{Baselines}

To evaluate the performance of {\tool}, we first identified the most relevant papers in code generation that utilized a common benchmark for comparison, i.e., 
papers that used \texttt{HumanEval} or \texttt{MBPP}, either exclusively or as part of their benchmarks. These two benchmarks were chosen due to their popularity in the field. 
Further narrowing down the selection, we applied three additional criteria: (1) whether their results were competitive (more than 90\% pass@1 on \texttt{HumanEval}), (2) whether their GitHub repository is available online, and (3) their cost is not disproportionately high. The third criterion was specifically defined to exclude SOTA tools with excessively high costs, ensuring that the selected tools remained competitive in terms of cost-effectiveness. For comparison, we used the \texttt{pass@1} metric. 
As a result, four papers are selected as our baselines, i.e., Reflexion~\cite{b26}, LATS~\cite{b27}, LDB~\cite{b29}, and AgentCoder~\cite{b28} (details are in Section~\ref{relatedWork}). For each baseline, we ran the experiments with the recommended configuration provided in its GitHub repository or provided in its paper. After the initial evaluation, AgentCoder~\cite{b28} was excluded from further analysis as we found that it is 6 times more expensive than the second most expensive alternative.
The baseline configurations used in our experiments are as follows:

\textbf{Reflexion}~\cite{b26}: We ran Reflexion with $max\_iters$ set to 2, $strategy$ set to ``reflexion'', $pass\_at\_k$ set to 1, and $model$ set to ``o3-mini''. Note that we have adapted Reflexion's code base to support the ``o3-mini'' model and tracking of token usage as these features were unavailable in Reflexion's original version.  

\textbf{LATS}~\cite{b27}: We ran the experiments with $max\_iters$ set to eight, $number\_of\_tests$ set to two, $strategy$ set to ``mctc'', $language$ set to ``py'', $pass@k$ set to 1, and $model$ set to ``o3-mini''. 
We also implemented the ``o3-mini'' model and tracking of token usage features in its codebase.  

\textbf{LDB}~\cite{b29}: We set $pass@k$ to $1$, $n\_proc$ to $1$, $max\_iters$ to $10$, $strategy$ to ``ldb'', and the model to ``o3-mini''. Consistent with previous works, we implemented the ``o3-mini'' model and tracking of token usage features as they were unavailable in the original version. 
The authors used Reflexion's generated solutions as seed programs to enhance their results. The seed programs served as the initial implementation of the code. We removed these seed files to ensure fairness, as other baselines do not use any seeds (an initial solution) to boost the performance. LDB uses the provided ground truth tests in the dataset for internal evaluation and feedback to the LLM. To enable a fairer comparison among the agents, we adopted a consistent test generation approach to produce test cases rather than using the ground-truth tests for internal evaluation. Using the original test cases for internal evaluation would be unfair to other baselines and impractical, as these test cases are meant solely for final evaluation, and are not typically available before code development. Consequently, we modified their test cases to use the generated test cases produced by the method employed in LATS and Reflexion (and ours).


\subsection{Experimental Setup}
\begin{table*}[t]
\centering
\caption{Default configuration parameters for EPiC}
\begin{tabular}{ |c|c|c|c|c|c }
\hline
 $mutation\_probability$ & $population\_size$ & $top\_n$ & $mutation\_alg$ & $temperature$ \\
\hline
 $0.4$ & $5$ & $10$ & $sim\_words\_as\_mutator$ & $0.0$\\ 
 \hline
\end{tabular}

\label{table:1}
\end{table*}

For the first RQ, we used HumanEval+, MBPP+, and BigCodeBenchHard. We used O3-mini, Sonnet 3.7, and Deepseek-v3 as the backend LLM for this RQ. For the second RQ, we used the same three datasets and O3-mini as the backend LLM. For the ablation study in Section \ref{RQ3}, we used BigCodeBenchHard dataset and o3-mini as the LLM. We used o3-mini for RQ2 and Section \ref{RQ3} as we found it to be the best-performing LLM in RQ1.

We used the default configuration setup for {\tool} presented in Table \ref{table:1}. The \texttt{Mutation\_probability} parameter specifies the likelihood of a word being substituted with a similar word. The \texttt{Population\_size} refers to the number of populations processed for each dataset instance, while \texttt{Top\_n} indicates the top number of related words considered for the mutation process. The \texttt{temperature} is the LLM temperature for code and test case generation. The defaults maintain a balance between exploration and stability. A mutation probability of 0.4 introduces sufficient variation while preserving the prompt's intent. A population size of 5 ensures diversity without being overwhelming, and considering the top 10 related words keeps substitutions semantically relevant. Using $sim\_words\_as\_mutator$ is cost-effective, and a temperature of 0.0 guarantees reliable results.


\section{Experiment Results}
\begin{table*}[t]
\footnotesize
    \caption{Performance comparison of various LLMs on HumanEval+, MBPP+, and BigCodeBenchHard datasets under Zero-shot and EPiC. The table presents pass@1 accuracy, token usage (in thousand), and associated costs for both Zero-shot and EPiC. The $\times$ symbol indicates the factor by which the token usage in EPiC exceeds that in the zero-shot setting.}
    \label{tab:performance_comparison}
    \centering
    \renewcommand{\arraystretch}{1.2}
    \begin{tabular}{l l c c c c c}
        \toprule
        \textbf{Dataset} & \textbf{LLM} & \textbf{Zero-shot} & \textbf{Zero-shot Token Usage}& \textbf{EPiC} & \textbf{EPiC Token Usage} & \textbf{EPiC Cost} \\
        & & (pass@1) &(thousand tokens)&(pass@1) & (thousand tokens) & (\$) \\
        \midrule
        \multirow{3}{*}{HumanEval+} 
        & o3-mini & 0.880 & \multirow{3}{*}{170} & 0.945 \textbf{(+6.5\%)} & 410 ($2.4\times$)  & 1.55 \\
        & claude 3.7 & 0.817 & &0.914 \textbf{(+9.7\%)} & 370 ($2.2\times$) & 3.74 \\
        & deepseek v3 & 0.866 & &0.932 \textbf{(+6.6\%)} & 540 ($3.2\times$) & 1.26 \\
        \midrule
        \multirow{3}{*}{MBPP+} 
        & o3-mini & 0.680 & \multirow{3}{*}{300} &0.743 \textbf{(+6.3\%)} & 1000  ($3.3\times$) & 4.03 \\
        & claude 3.7 & 0.674 & &0.724 \textbf{(+5\%)} & 800 ($2.7\times$) & 9 \\
        & deepseek v3 & 0.682 & &0.748 \textbf{(+6.6\%)} & 1040 ($3.5\times$) & 2.53 \\
        \midrule
        \multirow{3}{*}{BigCodeBenchHard} 
        & o3-mini & 0.350 & \multirow{3}{*}{200} &0.419 \textbf{(+6.9\%)} & 1660 ($8\times$) & 5.9 \\
        & claude 3.7 & 0.328 & &0.356 \textbf{(+2.8\%)} & 1180 ($6 \times$) & 9.98 \\
        & deepseek v3 & 0.338 & &0.356 \textbf{(+1.8\%)} & 1320 ($7 \times$) & 2.72 \\
        \bottomrule
    \end{tabular}
\end{table*}

\begin{table*}[t]
\footnotesize
\centering
\caption{Comparison of different approaches on HumanEval+, MBPP+, and BigCodeBenchHard datasets in terms of Pass@1 accuracy, token usage, and cost-effectiveness (ATSP). The best Pass@1 and ATSP values for each dataset are highlighted in bold.}
\label{tab:comparison}
\begin{tabular}{llcS[table-format=1.2]cc}
\toprule
Dataset & Method & {Token usage} & {Cost} & {Performance} & {Cost Effectiveness} \\
        &        &  {(thousand tokens)} &{(\$)}  &    {(Pass@1)}  & {(ATSP)} \\
\midrule
\multirow{4}{*}{HumanEval+}
 &  {\tool}       & 407  & 1.55  & \textbf{0.95} &\textbf{20} \\
 &  Reflexion  & 422  & 1.54  & 0.92         & 38 \\
& LATS       & 2429 & 9.30  & 0.93          &275 \\
&  LDB        & 489  & 1.43  & 0.89         &196 \\
& Zero-shot & 167 & 0.66 & 0.88 & -\\
\midrule
\multirow{4}{*}{MBPP+}
  & {\tool}       & 1003 & 4.03  & \textbf{0.74} & \textbf{31} \\
 &  Reflexion  & 970  & 3.48  & 0.72          & 44 \\
 & LATS       & 5614 & 17.58 & \textbf{0.74}  & 234 \\
 & LDB        & 1223 & 3.62  & 0.71          & 81 \\
 & Zero-shot & 298 & 1.25 & 0.68 & -\\
\midrule
\multirow{4}{*}{BigCodeBenchHard}
 & {\tool}      & 1666 & 5.90  & \textbf{0.42} & \textbf{141} \\
        & Reflexion & 1023 & 3.02  & 0.37          & 277 \\
        & LATS      & 6773 & 19.00 & 0.39          & 1110 \\
        & LDB       & 2204 & 5.90  & 0.38          & 450 \\
        & Zero-shot & 202 & 0.87 & 0.35 &-\\
\bottomrule
\end{tabular}
\end{table*}
\subsection{RQ1: How does {\tool} perform across different SOTA LLMs?}
In this RQ, we evaluate {\tool}’s performance in terms of functional correctness (measured by pass@1) and token usage across three benchmark datasets and using three SOTA LLMs (o3-mini, claude 3.7, and deepseek v3). The results, as summarized in Table \ref{tab:performance_comparison}, indicate that {\tool} consistently outperforms the baseline prompting approach across all settings while maintaining low costs.

For example, on the HumanEval+ dataset, pass@1 increases by approximately 6\%, 10\%, and 7\% for the respective LLMs, while token usage increases by a factor of 2.2 to 3.2 compared to zero-shot prompting. On the MBPP dataset, improvements in pass@1 range from 5\% to 7\%, with token usage rising by a factor of 2.7 to 3.5. In the case of BigCodeBench, pass@1 increases by 2\% to 7\%, and token usage increases by a factor of 6 to 8. Evidently, the more challenging the dataset, the greater the increase in token usage; for instance, HumanEval exhibits the lowest increase, whereas BigCodeBenchHard, the most challenging dataset, shows the highest token usage.

When comparing the LLMs, Claude 3.7 shows the lowest pass@1 and the highest cost, suggesting that it is less cost-effective than both o3-mini and deepseek v3. Meanwhile, o3-mini outperforms deepseek v3 on two datasets but incurs higher token costs. This pattern indicates that higher cost does not necessarily correlate with higher performance.

The results highlight a clear trade-off: {\tool} achieves higher functional correctness at the cost of increased token usage. However, this trade-off should be interpreted in the context of cost-effectiveness. Even though more tokens are used per instance, the incremental token consumption (and associated monetary cost) per additional problem solved remains low. For instance, despite o3-mini’s increased token usage on HumanEval+ (from 170 to 410 thousand tokens), the overall cost remains modest (\$1.55). Similar analyses for Claude 3.7 and deepseek v3 indicate that the additional cost is justified by the improved accuracy. Nonetheless, to fully assess the cost-effectiveness of {\tool}, it must be compared to SOTA code generation agents that employ iterative loops for code enhancement like {\tool}. In the next research question, we will compare the cost-effectiveness of {\tool} with these SOTA baselines.
\begin{tcolorbox}
    \textbf{Answer to RQ1:} {\tool} improves pass@1 across all LLMs and datasets while increasing token usage. Gains range from 2\% to 10\%, with token usage rising 2.2$\times$ to 8$\times$. o3-mini outperforms deepseek v3 but at higher costs, while Claude 3.7 is the least cost-effective. Despite increased token consumption, the accuracy gains justify the cost. 
\end{tcolorbox}




\subsection{RQ2: How does {\tool} compare to other iterative-based agents?}
\label{RQ2}

To evaluate {\tool}’s performance relative to Reflexion, LATS, and LDB, we compare pass@1 accuracy, token usage, and cost-effectiveness across three benchmark datasets: HumanEval+, MBPP+, and BigCodeBench (see Table \ref{tab:comparison}).


{\tool} consistently outperforms the baselines in terms of functional correctness across all datasets. In HumanEval+, {\tool} achieves a pass@1 of 0.95, surpassing Reflexion (0.92), LATS (0.93), and LDB (0.89). Similarly, in MBPP+, {\tool} attains a pass@1 of 0.74, slightly outperforming Reflexion (0.72), LATS (0.74), and LDB (0.71). In BigCodeBench, {\tool} achieves a pass@1 of 0.42, which is higher than Reflexion (0.37), LATS (0.39), and LDB (0.38). These results indicate that {\tool} consistently generates more functionally correct solutions compared to the iterative-based code generation agents. Overall, LATS ranks second, Reflexion third, and LDB fourth in terms of pass@1 accuracy.

While {\tool} demonstrates strong accuracy, it achieves this with a lower token usage than LATS and LDB, and it's comparable to Reflexion. For HumanEval+, {\tool} utilizes 407k tokens, which is less than LATS (2,429k) and LDB (490k), and only marginally less than Reflexion (422k). In MBPP+, {\tool} uses 1,003k tokens, while LATS consumes 5,614k tokens, Reflexion 970k, and LDB 1,223k. For BigCodeBench, {\tool}’s token usage is 1,666k, significantly lower than LATS (6,773k) and LDB (2,204k), but higher than Reflexion (1,023k). From a cost perspective, {\tool} is also highly efficient. In HumanEval+, {\tool} brings a total cost of \$1.55, similar to Reflexion (\$1.54) and lower than LATS (\$9.30). In MBPP+, {\tool} costs \$4.03, compared to Reflexion (\$3.48), LATS (\$17.58), and LDB (\$3.62). In BigCodeBench, EPiC maintains cost-effectiveness at \$5.90, on par with LDB but significantly lower than LATS (\$19.00).

The most critical metric, however, is cost-effectiveness, measured as Incremental Token Usage per Solved Problem (ATSP), as defined in Section \ref{metrics}. This metric accounts for both cost and effectiveness by comparing token usage and pass@1 accuracy relative to a zero-shot prompting baseline. {\tool} demonstrates the best cost-effectiveness (lowest ATSP score) across all datasets. For example, on HumanEval+, {\tool} consumes 20k extra tokens to find an additional correct solution relative to the zero-shot baseline, whereas Reflexion requires 38k, LATS 275k, and LDB 196k tokens. Thus, {\tool} is almost twice as cost-effective as Reflexion, 13 times more cost-effective than LATS, and 10 times more cost-effective than LDB. Reflexion ranks second, followed by LDB, with LATS being the least cost-effective.

A similar trend is observed for MBPP+ and BigCodeBenchHard. On MBPP+, {\tool} requires 31k tokens per additional correct solution, while Reflexion uses 44k, LATS 234k, and LDB 81k, making {\tool} 1.5 times more cost-effective than Reflexion, 7 times more cost-effective than LATS, and 2.5 times more cost-effective than LDB. On BigCodeBenchHard, {\tool}’s ATSP is 141k, whereas Reflexion, LATS, and LDB have ATSP values of 277k, 1,110k, and 450k, respectively, further demonstrating {\tool}’s better cost-effectiveness in the benchmark.

\begin{tcolorbox}
    \textbf{Answer to RQ2:} {\tool} outperforms Reflexion, LATS, and LDB in terms of functional correctness (pass@1) across all benchmark datasets. It achieves higher accuracy while maintaining lower token usage compared to LATS and LDB, and comparable effectiveness to Reflexion. In terms of cost-effectiveness, measured by Incremental Token Usage per Solved Problem (ATSP), {\tool} is the most cost-effective across all datasets, requiring significantly fewer tokens to generate additional correct solutions. Reflexion ranks second, followed by LDB, with LATS being the least cost-effective.
\end{tcolorbox}

\begin{table*}[t]
\footnotesize
\centering
\caption{Ablation study on the different mutation strategy and population size on {\tool} using BigCodeBenchHard dataset and o3-mini}
\label{tab:performance_metrics}
\begin{tabular}{l l c c c c}
\toprule
Row & \textbf{Mutator} & \textbf{Pass@1} & \textbf{Token Usage (thousand tokens)} & \textbf{Effectiveness (ATSP)} & \textbf{Population Size} \\
\midrule
1  & $\mathit{sim\_words\_as\_mutator}$ & 0.42 & 1666 & 141 & 5 \\
2 & $LLM\_as\_mutator$                   & 0.41 & 2457 & 253 & 5 \\
3 & $\mathit{sim\_words\_as\_mutator}$ & 0.37 & 2618 & 816 & 8 \\
\bottomrule
\end{tabular}
\end{table*}

\section{Discussion}
\label{RQ3}
 In this section, we conduct two concise ablation studies to compare the two mutation generator methods and to find the effect of changing the population size in {\tool}. \textit{LLM\_as\_mutator} utilizes an LLM in the mutation process, whereas \textit{sim\_words\_as\_mutator} employs locally calculated similarity matrices for mutation. The first method is more advanced but more costly too, while the second approach, significantly simpler, incurs minimal cost.  The first row in Table~\ref{tab:performance_metrics} uses the default \textit{sim\_words\_as\_mutator} and the second row uses \textit{$LLM\_as\_mutator$}. This finding suggests that a simple mutation tool is more cost-effective than an LLM-based tool, although this may not hold for all datasets. The third row uses a larger population size than the base in row one. By setting the population size to 8, we observe a drop in pass@1 and a decrease in effectiveness (higher ATSP value). This suggests that increasing population size beyond a certain threshold can make the results deviate from the optimal solution and unnecessarily inflates computational cost. These observations may change on different datasets or LLMs. One can use the preferred strategy on the desired dataset for optimal results. 

\section{Limitations and Threats to Validity}


\textbf{External:}
One external limitation of this research is that the datasets used (HumanEval+, MBPP+, and BigCodeBench) may not fully represent the diversity of real-world coding tasks, which may limit the generalizability of our findings. Another potential threat is that our findings, based on specific LLMs such as o3-mini, Sonnet 3.7, and DeepSeek-V3, may not generalize to LLMs from other vendors, such as Google's Gemini models. However, we chose LLMs from three different vendors to mitigate this threat.



\textbf{Construct:}
One construct threat we face is that our evaluation metrics rely on LLM-generated test cases for intermediate evaluations which could lead to misleading results if these test cases are not representative of the intended tasks. To ensure a fair evaluation, we deliberately use the same test generation process in both the baselines and {\tool}. Since the risk of invalid tests is present in all methods, the evaluation of the code generation process remains fair and valid. This limitation does not invalidate the results of the paper in any way.

\textbf{Internal:}
In this research, we compared our approach with baseline methods by using their publicly available code and default configurations, assuming they were optimized for best performance. For our implementation, we leveraged existing open-source packages for mutation to minimize internal validity concerns. However, we did not fine-tune our hyper-parameters. A fine-tuned {\tool} could potentially achieve even better performance than the baselines. Nevertheless, this does not pose any threats to the validity of our general findings.

\label{threats}

\section{Conclusion and Future Work}

In this paper, we introduced {\tool}, an evolutionary prompt engineering framework designed for cost-effective LLM-based code generation. Our approach leverages a lightweight evolutionary algorithm to optimize prompts, minimizing the number of interactions with LLMs while enhancing the functional correctness of generated code. Through extensive evaluations on benchmark datasets (HumanEval+, MBPP+, and BigCodeBenchHard), we demonstrated that {\tool} consistently outperforms SOTA methods in terms of cost-effectiveness, achieving a higher pass@1 score while consuming fewer tokens. Additionally, we introduced ITSP (Incremental Token usage per Solved Problem) as a new metric to quantify the cost-effectiveness for code generation, showing that {\tool} outperforms the baselines.

Despite its promising results, {\tool} opens several avenues for future research. First, while our approach has proven effective for code generation, it can be extended to other software applications, such as automated debugging, software documentation, and test case generation. Second, further exploration of multi-objective evolutionary algorithms could refine the prompt optimization process, simultaneously optimizing for accuracy, efficiency, and robustness. Additionally, integrating reinforcement learning could enhance {\tool}’s adaptability, enabling it to dynamically adjust mutation strategies based on real-time feedback. Finally, while our current implementation utilizes a predefined set of test cases, future work could explore self-improving evaluation methods that iteratively refine test case selection based on failure patterns observed during prompt evolution.

\bibliographystyle{ACM-Reference-Format}
\bibliography{sample-base}

\end{document}